\documentclass[amsmath, twocolumn, 10pt, aps, prb, superscriptaddress]%
	{revtex4-1}

\usepackage{xspace}
\usepackage{graphicx}

\newcommand{\dvec}[1]{\boldsymbol{#1}}
\newcommand{\vk}{\dvec{\mathrm{k}}}
\newcommand{\vq}{\dvec{\mathrm{q}}}
\renewcommand{\vr}{\dvec{\mathrm{r}}}

\newcommand{\mubar}{\ensuremath{\bar\mu}\xspace}
\newcommand{\dmu}{\ensuremath{\delta \mu}\xspace}

\newcommand{\nimp}{\ensuremath{n_{\mathrm{imp}}}\xspace}

\newcommand{\dmuloc}{\ensuremath{\delta \mu_{\mathrm{loc}}}\xspace}
\newcommand{\dmurms}{\ensuremath{\delta \mu_{\mathrm{rms}}}\xspace}

\newcommand{\mast}{\ensuremath{m^\ast}\xspace}

\newcommand{\cmsq}{\ensuremath{\mathrm{cm}^{-2}}\xspace}
\newcommand{\nm}{\ensuremath{\mathrm{nm}}\xspace}
\newcommand{\K}{\ensuremath{\mathrm{K}}\xspace}
\newcommand{\eV}{\ensuremath{\mathrm{eV}}\xspace}
\newcommand{\meV}{\ensuremath{\mathrm{meV}}\xspace}

\begin{document}

\author{D. S. L. Abergel}
\affiliation{Condensed Matter Theory Center, University of Maryland,
College Park, Maryland 20742, USA.}
\author{M. Rodriguez-Vega}
\affiliation{Department of Physics, College of William and Mary,
Williamsburg, Virginia 23187, USA}
\author{Enrico Rossi}
\affiliation{Department of Physics, College of William and Mary,
Williamsburg, Virginia 23187, USA}
\author{S. Das Sarma}
\affiliation{Condensed Matter Theory Center, University of Maryland,
College Park, Maryland 20742, USA.}

\title{Interlayer excitonic superfluidity in graphene}

\begin{abstract}
We discuss the conditions under which the predicted (but not yet
observed) zero-field interlayer excitonic condensation in double layer
graphene has a critical temperature high enough to allow detection.
Crucially, disorder arising from charged impurities and corrugation
in the lattice structure --- invariably present in all real samples ---
affects the formation of the condensate via the induced charge
inhomogeneity.  In the former case, we use a numerical
Thomas-Fermi-Dirac theory to
describe the local fluctuations in the electronic density in double
layer graphene devices and estimate the effect these realistic
fluctuations have on the formation of the condensate.
To make this estimate, we calculate the critical temperature for the
interlayer excitonic superfluid transition within the mean-field BCS
theory for both optimistic (unscreened) and conservative (statically
screened) approximations for the screening of the interlayer Coulomb
interaction.
We also estimate the effect of allowing dynamic contributions to the
interlayer screening.
We then conduct similar calculations for double quadratic bilayer
graphene, showing that the quadratic nature of the low-energy bands
produces pairing with critical temperature of the same order of
magnitude as the linear bands of double monolayer graphene. 
\end{abstract}

\maketitle

\section{Introduction}

The prediction of an interlayer direct Coulomb interaction driven
room-temperature excitonic condensate at zero magnetic field in double
monolayer graphene \cite{zhang-prb77,
min-prb78} (DMG), double quadratic bilayer graphene
\cite{perali-prl110} (DQBG), and hybrid monolayer-bilayer graphene
systems \cite{zhang-prl111} has captured much attention both for the
fundamental interest in the existence of a zero-field condensate and for
possible applications in devices, including ultra-fast switches and 
dispersionless field-effect transitors.  \cite{banerjee-ieeeedl30}
However, despite considerable experimental effort, the condensate has
yet to be observed in Coulomb drag experiments in zero magnetic field. 
\cite{gorbachev-natphys8, kim-prb83}
There are two possible reasons why this is so. First, 
the critical temperature ($T_c$) may simply be too small (i.e., much
lower than the optimistic mean-field theoretic predictions for ideal
systems) and therefore the condensate is destroyed by simple thermal
fluctuations. 
Second, it is possible that disorder in the form of electron--impurity
scattering, \cite{bistritzer-prl101, efimkin-jetplett93} scattering from
vacancy sites in the lattice, \cite{dellabetta-jpcm23} or from the
presence of inhomogeneity in the charge density distribution
\cite{abergel-prb86-b} is suppressing formation of the condensate.
In this work, we examine the last of these possibilities.
In order to avoid any confusion, we emphasize right at the beginning
that we are considering here only the case of interlayer superfluidity
in double-layer graphene systems and not the question of
superconductivity in individual (i.e., single layer) monolayer graphene 
and bilayer graphene, which is also an interesting problem and is
attracting a lot of attention.  
These two problems, i.e., the interlayer superfluidity of our interest
and superconductivity in individual graphene layers, are completely
distinct physical and mathematical problems. 
In the case of interlayer superfluidity, the individual layers of
graphene are not superconducting, only the interlayer Coulomb
correlations between the two layers drive the system into an interlayer
neutral superfluid.  Similar physics, often also referred to as
`excitonic condensation', could in principle occur in any two-component
(double-layer or two bands) electron-hole system if the inter-component
interaction is strong enough.

To address the issue of the critical temperature in the clean limit, 
it is known that the
size of the predicted excitonic gap in DMG depends very strongly on the
choice of screening of the interlayer Coulomb interaction.
\cite{zhang-prb77, min-prb78, sodemann-prb85, lozovik-prb86, 
kharitonov-prb78, lozovik-jetplett87}
For DMG embedded in a dielectric medium such as hBN or SiO$_2$, in the
situation where the two layers have equal chemical potential in opposite
bands, and where $k_F d = 0$ (where $k_F$ is the Fermi wave vector
and $d$ is the interlayer separation), it is
known that the prediction of room-temperature superfluidity is only
valid for the unscreened Coulomb interaction approximation.
\cite{sodemann-prb85}
Using static screening in the analysis (replacing the unscreened Coulomb
interaction) exponentially suppresses the predicted excitonic gap to a
regime where it is unmeasurable.
\cite{kharitonov-prb78, lozovik-jetplett87}
A dynamic screening approximation \cite{sodemann-prb85} 
gives a gap that is intermediate between these two. 
If self-consistent
screening of the Coulomb interaction including the gap at the Fermi
surface caused by the existence of the condensate is considered with
dynamic screening,\cite{sodemann-prb85} the superfluid gap sharply
increases if the effective interaction parameter $\alpha = e^2/(\epsilon
\hbar v_F) \gtrsim 1.5$ (or equivalently $\epsilon < 1.45$, where
$\epsilon$ is the dielectric constant of the medium and $v_F$ is the
Dirac band velocity of monolayer graphene). 
This self-consistent screening mechanism is also applied in the static
screening case \cite{lozovik-prb86} where the enhancement in the gap
size is found at $\alpha \approx 2.2$, corresponding to $\epsilon=0.9$,
implying all physical values of $\epsilon$ are above the threshold.
The impact of vertex corrections has also been studied
\cite{lozovik-prb86} and is claimed to be weak, although they may
slightly increase the pairing strength. 
Interband processes \cite{mink-prb84, sodemann-prb85} have also been
shown to slightly increase the critical temperature.
In the context of quadratically dispersing bands, room-temperature
superfluidity was predicted for electron--hole systems with ordinary
parabolic dispersion as early as 1976 by Lozovik and Yudson.
\cite{lozovik-jetp44}
The self-consistent screening method was also applied to DQBG
\cite{perali-prl110} where a similar enhancement of the pairing was
found at low carrier density (or, equivalently, small values of $k_F d$).
However, we believe there are some technical issues with the method of
the self-consistent screening calculation, which we explain and discuss
in the appendix -- these technical issues, although
important in their own right, are not particularly germane to the issue
of charge inhomogeneity, which is the main topic studied in our work.

The role of disorder has also been studied, and it was shown in
Refs.~\onlinecite{bistritzer-prl101, efimkin-jetplett93,
dellabetta-jpcm23} that intra-layer momentum scattering by short-range
disorder typically does not reduce $T_c$ substantially.
However, long-range disorder in the form of charge inhomogeneity does
play a key role \cite{abergel-prb86-b} because in this case the energy
scale to which the disorder must be compared is the excitonic gap,
which may be small depending on the choice of screening in the
interlayer interaction.
The carrier density inhomogeneities induced by the presence of
long-range disorder lead to differences in the chemical potential in the
two layers that remove the perfect nesting of the Fermi surfaces and
therefore strongly suppress $T_c$.
This effect is analogous to the Clogston-Chandrasekhar limit of
BCS superconductivity and the associated destruction of the pairing.
It must be emphasized that although the usual momentum scattering by
disorder does not affect the excitonic condensation by virtue of
Anderson's theorem, any density or chemical potential fluctuations
between the two layers would act as a random magnetic field for the
$s$-wave superconductor, strongly suppressing the superfluidity.  Random
charged impurities in the environment invariably lead to interlayer
chemical potential fluctuations because of their long-range Coulombic
nature.
One must therefore carefully distinguish between the disorder-induced
momentum scattering (which may not be particularly detrimental to the
predicted interlayer superfluidity) and the disorder-induced  density
inhomogeneity (which is extremely detrimental to the superfluidity).

The effects of density imbalance in the BCS-BEC crossover region have
also been investigated, \cite{pieri-prb75} but since the interlayer
interaction in graphene double layers is relatively weak, our results
are located firmly in the BCS regime and the added complexity of the
phase diagram near the crossover is not of immediate practical concern.

In this article, in Sec.~\ref{sec:monolayer}, we give a comprehensive
description of the conditions under which the condensate should be
observable in DMG. 
To do this, we accomplish two main tasks. First, we describe how $T_c$
behaves in the presence of a finite imbalance in the chemical potentials
of the two layers, a calculation which was not previously reported.
We estimate $T_c$ for both unscreened and statically screened
interlayer interactions as a function of the overall chemical potential
(\mubar) and asymmetry in the chemical potential in the two layers
(\dmu), and provide estimates for the dynamic screening case. 
Our reason for carrying out calculations using both unscreened and
statically screened Coulomb interactions is the fact that they
respectively represent the most optimistic and the most pessimistic
scenarios for the excitonic condensation to occur with the respective
$T_c$ estimates for the two approximations differing by several orders
of magnitude. 
Our theory thus provides upper and lower bounds on the expected $T_c$ in
the system without getting into the complications of attempting a
dynamical screening calculation of $T_c$ in the presence of disorder,
which is challenging and beyond the scope of the current work, and
remains a problem for the future.
The main aim is to describe the role of charge
inhomogeneity, which we can do within these two well-defined theoretical
approximations.
We find that any finite \dmu has the effect of reducing $T_c$, and that
when $\dmu$ is comparable to the size of the excitonic gap ($\Delta$),
$T_c$ becomes zero consistent with the Clogston-Chandrasekhar limit in
metallic superconductors.
The second task is to provide a comprehensive description of the nature 
of the charge inhomogeneity in DMG devices. We use Thomas-Fermi-Dirac 
theory (TFDT) \cite{rossi-prl101} to estimate the spatial fluctuation in
\dmu induced by randomly placed charged impurities. This is a completely
new application of TFDT and nothing similar has previously been
attempted to describe double layer graphene systems.
In this way, we gain a full understanding of the nature of the
correlations in the disorder of the two layers, and obtain
quantitatively accurate estimates of the spatial size and magnitude of
the charge fluctuations for realistic experimental parameters. 
We then link the two calculations by assuming that the variance of the
disorder-induced potential fluctuations represents the typical mismatch
or imbalance of the Fermi energies in the two layers. 
We find that when charged impurities are located close to
the DMG, the fluctuations in \dmu have a length scale of the order of 
$10\nm$, and that $\dmu \sim \mubar \gg \Delta$ indicating that the
excitonic condensate will not be able to form in this regime, as it is
suppressed totally by the disorder induced interlayer chemical
potential fluctuations.
However, when a clean spacer layer is used to separate the DMG from the
SiO$_2$ where the impurities reside, the fluctuations in $\dmu$ reduce
by an order of magnitude for comparable impurity densities. In this
regime, when the impurity density is low, we find that it is possible
for the excitonic condensate to form for reasonable $T_c$ estimates as
given by dynamic screening.

In Sec.~\ref{sec:bilayer} we present the condensate analysis of the DQBG
system at the same level of approximation as that of the DMG. This
analysis has also not been reported previously.
We find that the alteration in the single-particle band structure and
chirality properties of the underlying layers in going from DMG to DQBG
causes some qualitative change in the evolution of $T_c$ with \mubar
for the unscreened interaction, but that the gap size is quantitatively
similar for experimentally pertinent parameters. Crucially, the behavior
of $T_c$ with \dmu is unchanged.

As mentioned above, we believe that there are certain technical issues
with the method of the self-consistent screening calculations presented
in other works \cite{perali-prl110, sodemann-prb85, lozovik-prb86} and
we discuss this in Appendix \ref{app:screening}.

\section{Double monolayer graphene \label{sec:monolayer}}

\subsection{$T_c$ for asymmetrically doped layers}

\begin{figure}[tb]
	\centering
	\includegraphics[]{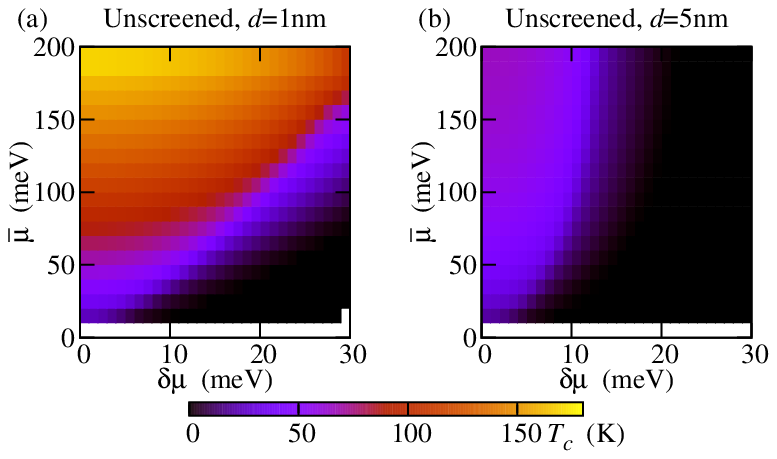}\\%
	\includegraphics[]{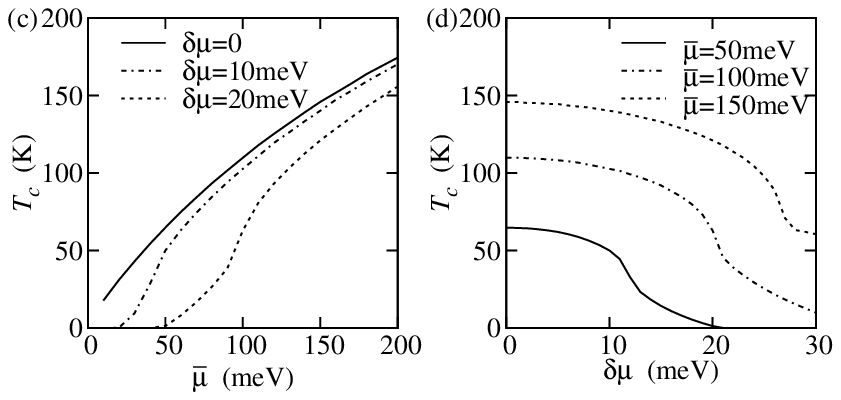}
	\caption{(Color online.)
	$T_c$ for the unscreened interaction in dielectric environment
	$\epsilon=3.9$. (a) and (b) show color plots as a function of
	$\mubar$ and $\dmu$ for $d=1\nm$ and $d=5\nm$, respectively. (c)
	$T_c$ as a function of $\mubar$ for various $\dmu$ and $d=1\nm$. (d)
	$T_c$ as a function of $\dmu$ for various $\mubar$ and $d=1\nm$.
	\label{fig:Tc}}
\end{figure}

We begin by describing the role of a finite chemical potential
difference between the two layers in reducing $T_c$.
A rigorous estimate of $T_c$ in the non-disordered case is already an
intractable calculation since the inclusion of dynamic screening effects
in the interlayer interaction demands that the vertex corrections be
included in the theory.  In fact, there is no
Migdal's theorem when considering superconductivity (or fermionic
superfluidity) induced by electron-electron interactions, and thus vertex
corrections must, in principle, be included in the theory even for the
unscreened interaction. As such, all mean-field BCS type theories of
interlayer excitonic superfluidity are somewhat suspect strictly from a
theoretical viewpoint since vertex corrections are uncritically
neglected.  However, in the presence of
dynamic screening, the theory becomes particularly suspect if vertex
corrections from the ladder diagrams, which contribute to Migdal's
theorem, are left out since the interaction itself now is calculated
in an approximation including infinite number of electron-hole loops.
The role of vertex corrections and other processes which go beyond
Migdal's theorem have been discussed in
Ref.~\onlinecite{grimaldi-prl75}.
Therefore, since all analytically approachable calculations will either
over-estimate (i.e., the unscreened case) or under-estimate (i.e., the
statically screened case), the actual $T_c$, we shall
show how the chemical potential imbalance will affect the formation of
the excitonic condensate for two distinct model calculations. 
We consider our calculation not to give a quantitative estimate of the
real $T_c$, but to provide a comprehensive survey of the role of
disorder in the form of charge inhomogeneity.
Using the
unscreened interlayer interaction (which will systematically
over-estimate $T_c$), and the statically screened interlayer
interaction (which will under-estimate $T_c$), we show that when
$\dmu > 0$ is less than the excitonic gap (which we label $\Delta$), 
$T_c$ is reduced but remains finite. When $\dmu \sim \Delta$, $T_c$
becomes zero.
Thus, the disorder-induced chemical potential imbalance or asymmetry is
a key parameter determining the existence or absence of the excitonic
condensation, which needs to be taken into account in the experimental
search for interlayer superfluidity.
 
\begin{figure}[tb]
	\centering
	\includegraphics[]{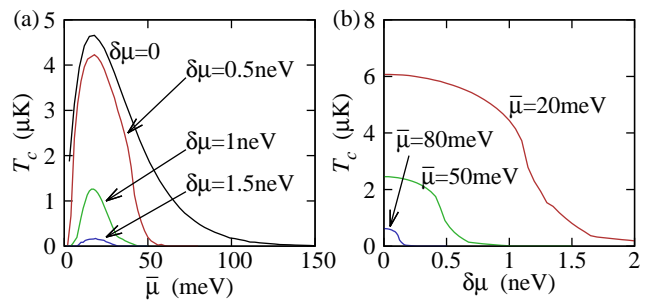}
	\caption{(Color online.) Static interlayer screening for DMG with 
	$\Delta_{\vk} = \Delta_{k_F}$. (a) For various \dmu as a function of
	\mubar, and (b) for various \mubar as a function of \dmu.
	Note that the scale on the vertical
	axis is micro-Kelvin.
	\label{fig:monstatic}}
\end{figure}

It is convenient to assume that the upper layer is doped with
electrons, the lower layer with holes, and to perform a particle-hole
transformation in the lower layer such that both layers are described by
a chemical potential with positive sign. We then characterize the system
by their average chemical potential $\mubar = (\mu_u + \mu_l)/2$ and
difference $\dmu = \mu_u - \mu_l$. 
This implies that, for a given \mubar, the maximum value of \dmu
possible is $2\mubar$. 
Note that in other works relating to the excitonic condensate in DMG, it
was assumed that $\mu_u = \mu_l$
giving perfect nesting of the Fermi surfaces in the two layers.
\cite{zhang-prb77, min-prb78, perali-prl110, bistritzer-prl101,
efimkin-jetplett93, dellabetta-jpcm23, sodemann-prb85, kharitonov-prb78,
lozovik-jetplett87, lozovik-prb86, mink-prb84, lozovik-jetp44}
Within the mean-field theory and the BCS approximation, the
temperature-dependent gap function is given by\cite{abergel-prb86-b}
\begin{equation}
	\Delta_{\vk}(T) = \sum_{\vk'} V(\vk'-\vk) \frac{\Delta_{\vk'}(T)
	f(\vk,\vk') N(\vk',T)}{E_{\vk'}}.
	\label{eq:Delta}
\end{equation}
In this equation, $V(\vq) = V(\vq,0)$ is the static limit of the random
phase approximation (RPA) for the interlayer potential,
$f(\vk,\vk') = [1 + \cos(\theta_k - \theta_{k'})]/2$ comes from
the chirality of electrons in monolayer graphene,
$N(\vk',T) = n_\beta(\vk',T) - n_\alpha(\vk',T)$ is the 
finite temperature occupation factor of the excitonic bands labeled by
$\alpha$ and $\beta$, which contain $\delta \mu$ and $E_{\vk'} =
\left[(v_Fk'-\mubar)^2 +
\Delta_{\vk'}^2\right]^{1/2}$, where $v_F$ is the monolayer graphene
Fermi velocity. We identify the excitonic gap $\Delta$ 
as the peak value of this function, which, in the BCS limit, is found at
$k=k_F$.
We define $T_c$ as the lowest value of $T$ for
which the gap function is zero for all $k$. We find this condition
numerically, and it gives a value which is of the same order of the
standard prediction $\Delta = 1.76 T_c$ from the textbook constant gap
approximation to mean-field theory. 
Our value for $T_c$ is a little smaller due to the non-constant gap
function and full momentum-dependent interaction, which we retain in the
calculation.
We note as an aside that in the presence of the Fermi surface mismatch
in the two layers (i.e., a chemical potential or density imbalance
between the layers), there can, in principle, be inhomogeneous FFLO type
solutions for the ground-state superfluidity in the system, but our
general calculations allowing for the possibility of such inhomogeneous
FFLO states fail to find any FFLO solutions for either the DMG or the
DQBG systems, and we consistently find either purely homogeneous
superfluid condensate or no condensate.

The interlayer screened interaction potential is calculated within the
RPA, which is justfied for double layer graphene because the fermion
number $N=8$ is large, as
\begin{equation}
	V(\vq,\omega) = \frac{v_q e^{-qd}}%
	{1 + 2 v_q\left(\Pi_u + \Pi_l\right)%
	+ v_q^2 \Pi_u \Pi_l(1 - e^{-2qd})},
	\label{eq:Vqomega}
\end{equation}
where $v_q = 2\pi e^2 /(\epsilon q)$, 
$\Pi_u$ and $\Pi_l$ are the polarization functions for the upper
and lower layers, respectively, and are functions of $\vq$ and
$\omega$. The polarization functions are given for both the static
($\omega = 0$) limit and dynamic approximations in
Ref.~\onlinecite{hwang-prb75}.
The unscreened and static screening cases represent the two limits for
the size of the gap in a real system: the unscreened interaction tends
to overestimate the pairing and hence gives a large $T_c$, while the
static screening tends to overestimate the screening efficiency and
therefore yields a small $T_c$. 
In this article, we consider both these cases making the reasonable
assumption that the
reality of the situation is somewhere in-between. \cite{sodemann-prb85}

In the optimistic case of the unscreened interlayer Coulomb
interaction, close layer separation ($d=1\nm$), and a flat hBN substrate
with $\epsilon=3.9$, we find that $T_c$ can be of the order of 100 K for
realistic doping and moderate layer imbalance [see
Fig.~\ref{fig:Tc}(a)]. 
As clarified in Fig.~\ref{fig:Tc}(c), $T_c$ is a monotonically
increasing function of $\mubar$, and larger \dmu reduces $T_c$.
Figure \ref{fig:Tc}(d) shows that as the chemical potential asymmetry
increases, $T_c$ decreases monotonically with a characteristic ``S''
shape.
Note that our theory neglects the possibility of direct tunneling
between the layers, but this may become a significant factor and could
even enhance the formation of the condensate \cite{mink-prb84} as the
interlayer spacing becomes this small.
Taking $d=5\nm$, which is thick enough to fully suppress interlayer
tunelling [see Fig.~\ref{fig:Tc}(b)], gives $T_c \approx 50\K$. 
We show our results in terms of \mubar and \dmu because these are the
intuitive variables in the theory and allow for straightforward
comparison with the results of the inhomogeneous case in
Sec.~\ref{sec:inhomog}. However, for experimental comparison, it is more
convenient to parametrize in terms of carrier density than chemical
potential. Straightforwardly, in the single-particle case,
we have $\bar{n} = \mubar^2/(\pi\hbar^2
v_F^2)$, but $\delta n$ depends non-linearly on the fluctuation and the
chemical potential as $\delta n = \dmu(2\mubar + \dmu)/(\pi \hbar^2
v_F^2)$.

We have repeated these calculations using the statically screened
interlayer interaction finding that the gap is so small that our
numerical procedure for $T_c$ cannot resolve it within acceptable error
bars. Therefore, we suggest that the gap in this case is essentially
zero even for $d=1\nm$ and $\epsilon=1$.
This happens because $T_c$ is so small that the occupation factors, which
appear in the self-consistent gap equation are very steep functions near
the Fermi energy. Therefore, in order to produce a non-zero numerical
result for $\Delta$ in this case, we approximated the gap as momentum
independent.
This is a reasonable approximation, also used, for example, in
Ref.~\onlinecite{kharitonov-prb78}.
Doing this yields the results shown in Fig.~\ref{fig:monstatic}, which
can be taken as a reliable upper bound on $T_c$ with static screening. 
Figure \ref{fig:monstatic}(a) shows $T_c$ as a function
of \mubar for several different values of \dmu, $\epsilon=3.9$, and
$d=1\nm$. The behavior of $T_c$ is qualitatively different from the
unscreened case in that there is a maximal value of $T_c$, which depends
sensitively on both $\epsilon$ and $d$. This is because as the carrier
density increases, the screening becomes more effective and the
interlayer interaction is reduced. To illustrate the dependence
of $T_c$ on \dmu, in Fig.~\ref{fig:monstatic}(b) we show this for a few
different values of \mubar. Notice that the scale on the horizontal axis
is $10^{-9}\eV$, indicating that a tiny layer imbalance is enough to
kill the condensate. 
However, the qualitative behavior of $T_c$ with
\dmu is identical to the unscreened case.

\begin{figure}[tb]
	\centering
	\includegraphics[]{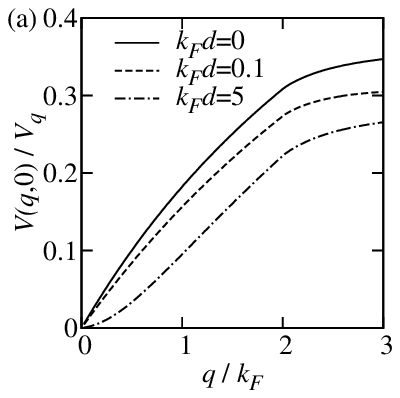}%
	\includegraphics[]{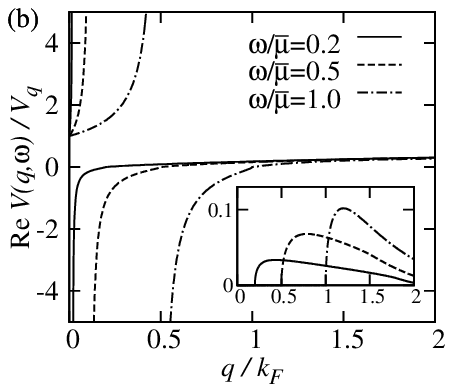}
	\caption{ Comparison of (a) the statically screened and (b) the real
	part of the dynamically screened interlayer interaction potentials
	to the unscreened interaction $V_q=2\pi^2e^{-qd}/(\epsilon q)$ for
	DMG with $\epsilon = 3.9$. The inset to (b) shows the imaginary part
	of the interaction.
	\label{fig:VdynComp}}
\end{figure}

To demonstrate why the statically screened interaction gives a $T_c$
that is so low, in Fig.~\ref{fig:VdynComp}(a) we show as a function
of momentum $q$ the statically
screened interaction normalized by the unscreened potential $V_q = v_q
e^{-qd}$. The potential is a universal function of $q/k_F$ for $d=0$,
but is weakly dependent on $k_F d$ for finite $d$. We show representative
curves for $k_F d=0$, $k_F d\ll 1$, and $k_F d>1$ corresponding to the
strongest interaction limit, the low-density limit, and the high-density
limit, respectively. We see that higher carrier density reduces the
interaction strength since the increased density of states allows the
screening to be more efficient.
The static polarizability is constant up to $q=2k_F$, but increases
after that,\cite{hwang-prb75} causing the noticeable flattening of the
interaction potential as a function of $q$ in
Fig.~\ref{fig:VdynComp}(a).
The peak of the gap function is found at $k=k_F$ and the most relevant
contribution to the integrand in Eq.~\eqref{eq:Delta} comes from $k = k'
\approx k_F$ indicating that the important range of wave vectors is
$|q|<k_F$. 
In this regime, the statically screened potential is an order
of magnitude smaller than the unscreened one leading to an excitonic
gap, which is several orders of magnitude smaller than the unscreened
case, consistent with previous analytical evaluations.
\cite{kharitonov-prb78}

The dynamically screened  form of the interaction has been used by other
authors, so we analyze the potential function \eqref{eq:Vqomega}
itself to gain some intuitive insight into the effect of this
approximation.
In Fig.~\ref{fig:VdynComp}(b), we show the dynamically screened
interaction potential normalized by $V_q$ as a function of wave vector
for various frequencies and $k_Fd = 0$. In this limit, the interaction
is a universal function of $q/k_F$.
The dynamic screening is a strong function of the frequency $\omega$,
however, some instructive patterns can be identified. 
For $q/k_F < \omega/\mubar$, the polarization functions are negative
\cite{hwang-prb75} and therefore it is possible for the finite
frequency potential to have a divergence corresponding to the plasmon
wave vector $k_p$. 
The potential also is negative \footnote{Note that the overall negative
sign for an attractive interaction is absorbed into the definition of
the gap function in Eq.~\eqref{eq:Delta}.} for $k_p/k_F < q/k_F <
\omega/\mubar$ indicating an overall repulsive interaction which could
reduce the gap size.
In contrast, we note that for $q < k_p$, the interaction is
enhanced over the unscreened case. 
This is the well-known anti-screening effect of dynamic screening,
which should to some extent compensate for the sign change of the
interaction in some regime of the phase space.
For $q/k_F > \omega/\mubar$, the potential is very similar to the
statically screened case. 
The most relevant frequency range is $\omega \approx \Delta$, which is
in general rather small compared to $\mubar$ and hence the range of wave
vectors where the potential deviates substantially from the statically
screened case is small, indicating that $T_c$ with dynamic screening
will be closer to that predicted by the static screening calculation
than the unscreened one.
This analysis also shows that a system with a large bare gap will be
more robust against dynamic screening effects since the relevant
frequency will be higher, implying that within the range of $q$ that
contributes strongly to the integrand in Eq.~\eqref{eq:Delta}, we have
$V(q,\omega)/V_q > 1$.

Calculations for the unscreened interaction for suspended DMG with
$\epsilon=1$ and $d=1\nm$ show that the excitonic gap is large
with respect to $\mubar$ and therefore fluctuations of the order of the
chemical potential will not reduce $T_c$ to zero.
When $d=5\nm$, we find that $T_c$ drops to room temperature or a little
below. 
Suspended graphene is also known to form ripples with size fluctuations
of the order of $1\nm$ in height,\cite{zan-nanoscale4} which may make
the precise control of the interlayer spacing difficult for these
proposed devices, and which may introduce charge inhomogeneity related
to the strain field induced by the corrugations.
We shall discuss the effects of ripples in the next section, after we
have described the role of charge inhomogeneity in DMG.

\subsection{Charge inhomogeneity in DMG \label{sec:inhomog}}

In any experimental sample, some degree of extrinsic disorder-induced 
charge inhomogeneity will
exist, as has been demonstrated by many surface measurements of
monolayer graphene. \cite{martin-natphys4, decker-nl11,
xue-natmat10}
In a double layer device, the inhomogeneities in the charge landscapes
will not be identical in both layers, and therefore there will be
spatial variation in the asymmetry of the chemical potentials.
In this situation, the local difference in chemical potential has two
contributions. There is a nominally homogeneous part which is induced by
gating and is, in principle, controllable. This contribution was the
subject of the previous section and we ignore it here. 
Then there is a contribution from charged impurities and other disorder
that is inhomogeneous and not controllable.
A full analytical description of the inhomogenous system is clearly
intractable so we employ an accurate numerical method to compute the
charge density of the system when charged impurities explicitly break
translational symmetry.
From this charge landscape, we can assign the local chemical potential
$\mu_u(\vr)$ and $\mu_l(\vr)$ in each layer, and characterize the
spatial fluctuations by their root-mean square (rms) value, which is a
measure of the typical fluctuation.
Using this measure of the disorder in the charge landscape, we can
discuss the stability of the condensate against the density and chemical
potential inhomogeneity induced by the charged impurities.
In principle, it is possible that some correlation will exist between
charged impurities, although the nature of these correlations will
depend on details of the system. We wish to avoid introducing extra
parameters to describe this, so we assume uncorrelated disorder for the
purposes of this work. If correlations are shown to be important, then
they can be included within the theory we are about to describe in the
same way as Ref.~\onlinecite{li-prl107}.

\begin{figure}[tb]
	\centering
	\includegraphics[]{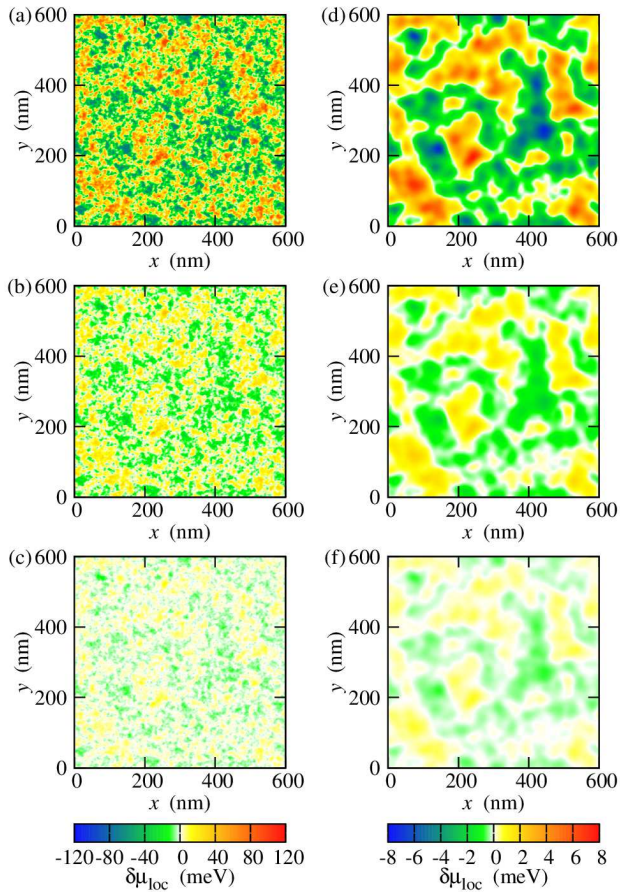}
	\caption{(Color online.) 
	Spatial plots of \dmu calculated via the TFDT. 
	The left column is for $d=1\nm$, $d_B=1\nm$, and $\mubar=50\meV$,
	which roughly approximates the experimental situation in 
	Ref.~\onlinecite{kim-prb83}.
	The right column is for $d=5\nm$, $d_B=20\nm$, and
	$\mubar=200\meV$,
	which roughly approximates the experimental situation in 
	Ref.~\onlinecite{gorbachev-natphys8}.
	The color bar at the bottom of each column applies to all three
	plots in each column. The first row is for $\nimp = 10^{11}\cmsq$, the
	second row is $\nimp=10^{10}\cmsq$, the third row is $\nimp =
	10^{9}\cmsq$.
	\label{fig:spatialdmu}}
\end{figure}

To calculate the charge landscape in each layer taking into account the
presence of long-range disorder due to the charged impurities
and nonlinear screening effects we use the TFDT. \cite{rossi-prl101}
The TFDT is a generalization to Dirac materials of the Thomas-Fermi
theory. \cite{giuliani-2005} TFDT and density-functional-theory
(DFT) \cite{hohenberg-pr136, kohn-pr140, kohn-rmp71, polini-prb78}
are similar in spirit in the sense that they are both non-perturbative
functional theories and therefore have the great advantage of being able
to take into account non-linear screening effects that are dominant in
systems like graphene at low doping. 
The key difference between TFDT and DFT in that in the latter, the
kinetic energy operator is retained, while in the former it is replaced
with a density functional. 
As a consequence, TFDT is computationally much more efficient than DFT
and can be used to study large systems in the presence of long-range
disorder where DFT is completely impractical. 
In particular, by using the TFDT, we are
able to obtain disorder-averaged results which would be impossible with
the strict DFT approach because of the heavy numerical cost. The
simplified treatment of the kinetic energy term limits the validity of
TFDT to regimes in which $|\nabla n/n|<k_F$.  
We have verified that this condition is reasonably well satisfied in 
single-layer graphene. \cite{rossi-prl101, rossi-prb79}
Our results show that in graphene double layers, due to the increased
screening of the disorder potential caused by the presence of an
additional graphene layer, the correlation length of the
disorder-induced inhomogeneities is larger than in isolated single-layer
graphene and therefore that the condition $|\nabla n/n|<k_F$ is always
well satisfied in these double layer graphene heterostructures.

The carrier density in the ground state is obtained by minimizing the
TFDT energy functional
\begin{multline}
 E[n_u,n_l] = E_u[n_u(\vr)] + E_l[n_l(\vr)] \\ 
	+ \frac{e^2}{2\epsilon} \int d^2\vr \int d^2\vr'
	\frac{n_u(\vr)n_l(\vr')}{(|\vr-\vr'|^2 + d^2)^{1/2}}
	\label{eq:tfdt}
\end{multline}
where $E_i[n_i(\vr)]$ is the energy functional of the density profile
for the $i$-th graphene layer (as given in Ref.~\onlinecite{rossi-prl101})
and the last term is the interlayer Coulomb interaction. 
Each layer functional $E_i[n_i(\vr)]$ contains a term due to the
disorder potential $V_D$ created by the charged impurities. We also
include intra-layer exchange interactions.
We assume that charged impurities located close to the surface of the 
SiO$_2$ constitute the dominant source of disorder and we therefore
model the charged impurity distribution as an effective two-dimensional
distribution $C(\vr)$ placed at a distance $d_B$ below the lower graphene
layer.
Note that this is the most generous estimate for the charged impurities
that we can take, since we are neglecting any disorder at the other
interfaces which may be induced by the successive fabrication steps
required to make these devices.\cite{dean-ssc152}
Denoting disorder-averaged quantities by angle brackets, without loss of
generality, we assume $\langle C(\vr)\rangle=0$.  We also assume the
charged impurities to be uncorrelated \cite{rossi-prl101, li-prl107,
li-ssc152} so that $\langle C(\vr) C(\vr')\rangle =
\nimp\delta(\vr-\vr')$, where $\nimp$ is the charged impurity density.
$V_D$ in each layer is the Coulomb potential created by the random
distribution $C(\vr)$. 
The ground-state density distributions $n_u(\vr)$ and $n_l(\vr)$ are
obtained by minimizing $E[n_u,n_l]$ numerically enforcing the
self-consistency of the distribution in the two layers due to the
interlayer interaction. 
Then, the local difference in chemical potential between the two layers
$\dmuloc=\mu_u(\vr) - \mu_l(\vr)$ can be extracted for each point in the
system and by performing the minimization for many ($\sim 600$)
disorder realizations  and we obtain statistics for the distribution
function of \dmuloc.

\begin{figure}[tb]
	\centering
	\includegraphics[]{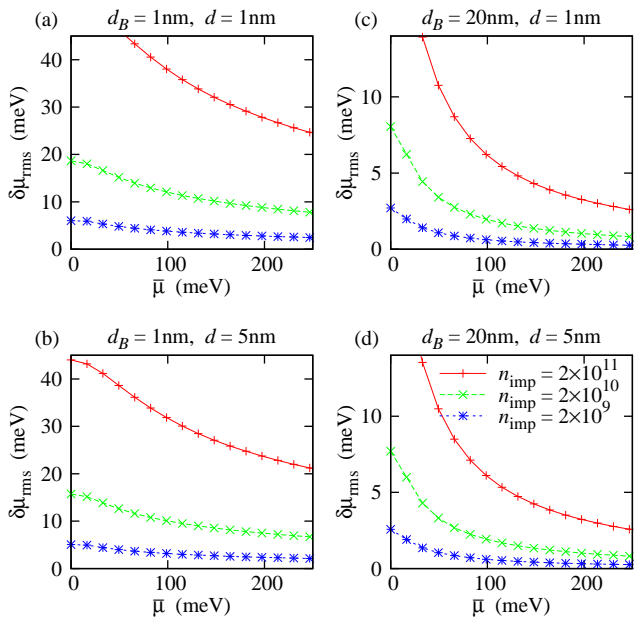}
	\caption{(Color online.) 
	Root-mean-square of the distribution of the local 
	$\delta \mu$ as a function of the global chemical potential for two
	experimentally relevant geometries.
	(a),(b) $d_B=1\nm$, $\epsilon=3.9$ corresponding to double layer
	graphene placed straight onto a SiO$_2$ substrate per the
	experiments in Ref.~\onlinecite{kim-prb83}.
	(c),(d) $d_B=20\nm$, $\epsilon=3.9$ corresponding to double
	layer graphene placed onto a 20$\nm$ slab of hBN per the
	experiments in Ref.~\onlinecite{gorbachev-natphys8}.
	\label{fig:avdmu}}
\end{figure}

In Fig.~\ref{fig:spatialdmu} we show the spatial distribution of \dmuloc
for three different impurity densities and in two experimentally
relevant geometries. 
In the left column, we show data where the lower graphene layer is
placed directly onto an SiO$_2$ substrate and the interlayer separation is
$d=1\nm$. 
In this case, the charged impurities at the oxide interface are
effectively approximated by a two-dimensional distribution 1\nm below
the lower graphene layer, so we take $d_B=1\nm$.
This corresponds to the system used in the experiments in
Ref.~\onlinecite{kim-prb83}.
We set $\mubar=50\meV$ corresponding to an easily achievable carrier
density regime.
The right column corresponds to a system where the lower graphene
layer is seperated from the SiO$_2$ substrate by a $20\nm$ layer of
hBN, like that used in the experiments in 
Ref.~\onlinecite{gorbachev-natphys8}.
Hence we take $d_B=20\nm$, and we also set $\mubar=200\meV$
corresponding to the high carrier density regime where we expect the
screening of the external impurities to be the most efficient, resulting
in the lowest amount of charge inhomogeneity.
In both situations we assume that the gate-induced (homogeneous,
controllable) part of the layer asymmetry to be zero. 
Therefore we assume that any layer imbalance is completely defined by
the charged impurities.
The rows of Fig.~\ref{fig:spatialdmu} show the data for, from top to
bottom, $\nimp=10^{11}\cmsq$, $10^{10}\cmsq$, and $10^9\cmsq$.
(We note that the higher value of \nimp is more typical, and
$\nimp=10^9\cmsq$ is unlikely to be achieved in laboratory graphene
samples on any substrate. Typically, one can get an estimate of \nimp
in a particular sample by looking at the carrier density regime over
which the graphene minimum conductivity ``plateau'' exists around the
Dirac point. \cite{dassarma-rmp83})
All six plots share some similar qualitative features. 
In particular, they all show regions where \dmu is positive and regions
where it is negative
with narrow strips in between where \dmu is small.\footnote{It is
possible that these narrow regions with $\dmu \approx 0$ will allow the
formation of percolating stripe states, but since our analysis of the
BCS theory is not valid in this highly inhomogeneous regime, we do not
examine it further.} The lengthscale of the fluctuations is not affected
by the impurity density, but the magnitude of the fluctuations is.
By comparing the data in the two columns we see that the distance of the
impurities from the DMG makes a substantial difference in the length
scale of the fluctuations in \dmuloc, and in reducing the magnitude of
the fluctuations.
The density of impurities also has a strong effect on the magnitude of
the fluctuations in \dmuloc with the fluctuations reducing by
approximately a factor of three with each order of magnitude decrease in
\nimp.
In the most dirty case (it should be noted that graphene on SiO$_2$ can
have an impurity density of up to $5\times 10^{12}\cmsq$ as measured by
transport measurements\cite{dassarma-rmp83}), 
shown in Fig.~\ref{fig:spatialdmu}(a) 
then the fluctutations in \dmuloc may be of the order of $\mubar$,
indicating that the condensate has no opportunity to form in this case.
For the cleanest situation shown in Fig.~\ref{fig:spatialdmu}(f), the
potential imbalance is on the scale of 1\meV and there is a significant
chance that excitons with a gap of the size predicted by dynamic
screening calculations \cite{sodemann-prb85} will persist in spite of
the disorder if an impurity density as low as $10^9\cmsq$ can be
achieved in double layer graphene samples.

We take many disorder realizations ($\sim 600$) for each impurity
density and collect ensemble-averaged statistics for the distribution of
\dmuloc. 
We characterize this distribution by its root-mean-square value, which
we label \dmurms.
In Fig.~\ref{fig:avdmu} we plot \dmurms for the two
experimental geometries discussed above and for different interlayer
spacing $d$. 
This is shown in Figs.~\ref{fig:avdmu}(a) and \ref{fig:avdmu}(b) 
for $d=1\nm$ and $d=5\nm$,
respectively, for three impurity densities covering three orders of
magnitude. The fluctuations are strongest at low carrier density, where
the screening of the impurity potential is weakest, and it decreases
monotonically with increasing \mubar. The trend suggested by the spatial
plots is confirmed here, that is, decreasing the impurity density by a
factor of ten generates approximately a factor of three reduction in the
fluctuations.
If the impurities are moved away from the DMG by a spacer layer as in
Figs.~\ref{fig:avdmu}(c) and \ref{fig:avdmu}(d) we find that the
fluctuations in \dmu are reduced to the order of 1\meV. 
This degree of fluctuation may be small enough to allow the condensate
to be detected at a reasonable temperature scale.
For the $d_B=1\nm$ case, a system dimension of 160 unit cells was
chosen. For $d_B=20\nm$ we increased that to 320 unit cells to ensure
that the system was large enough to accurately capture the size of the
typical fluctuations.

We briefly discuss the role of corrugations and ripples in the structure
of the graphene lattice and describe their effect on the excitonic
condensate. 
The existence of these ripples in suspended \cite{zan-nanoscale4}
samples and those placed on substrates of different kinds
\cite{xue-natmat10} have been demonstrated.
In the case of suspended monolayer graphene, ripples
of height of several angstroms have been observed,\cite{zan-nanoscale4} 
indicating that a suspended double layer structure with small
interlayer spacing may be difficult to control since ripple
corrugations will then be of the same size as the layer separation.
It has been theoretically predicted 
\cite{gibertini-prb81, partovi-azar-prb83, gibertini-prb85}
that the strain field associated with ripples can give rise to
fluctuations in the local density of the order of $10^{12}\cmsq$. 
If this is correct then it shows that lattice corrugations may
be a significant barrier to the existence of the condensate, since this
will give $\delta \mu \approx 40\meV$ which would destroy the condensate
even for unscreened Coulomb interactions.

\section{Double bilayer graphene \label{sec:bilayer}}

We now consider the analogous situation for double quadratic bilayer
graphene (DQBG). 
In this case, two AB-stacked bilayer graphene sheets \cite{mccann-prl96}
replace the monolayers discussed previously. 
We employ the same approach as we did for DMG in that we analyze the
critical temperature of the excitonic superfluid for the unscreened and
statically screened interlayer interactions, and then examine the
modification of the interlayer potential with the inclusion of dynamic
screening effects.
We model the two bilayer graphene sheets as having a gapless, quadratic 
low energy band structure $E_{\nu k} = \nu \hbar^2 k^2/(2\mast)$ where
\mast is the effective mass which we assume to be the same in both
layers, and $\nu = \pm1$ denotes the band.
Note that this approximation is only valid at densities
$n < 3\times 10^{12}\cmsq$, which is approximately equivalent to $k_F =
0.3\nm^{-1}$ and $\mu \approx 70\meV$.
The difference in the low-energy band structure leads to qualitative
changes in the behavior of the condensate as a function of \mubar, but
not of \dmu, as we shall demonstrate.
The gap equation in Eq.~\eqref{eq:Delta} and interlayer interaction in
Eq.~\eqref{eq:Vqomega} are still valid in the double bilayer case,
except for a redefined angular factor $f(\vk,\vk') = \cos^2(\theta_k -
\theta_{k'})/4$, exciton energy $E_{\vk'} = \left[ (\hbar^2
k'^2/(2\mast) - \mubar)^2 + \Delta_{\vk'}^2 \right]^{1/2}$, and the
polarization screening function of the quadratic bilayer obviously needs
to be taken into account.\cite{sensarma-prb82}
Note that this is a similar system to that studied by quantum Monte
Carlo methods recently,\cite{maezono-prl110} where continuous
transitions between a one-component fluid phase, an excitonic fluid
phase, and a biexcitonic phase were predicted. However, that publication
did not include any form of disorder.

\begin{figure}[tb]
	\centering
	\includegraphics[]{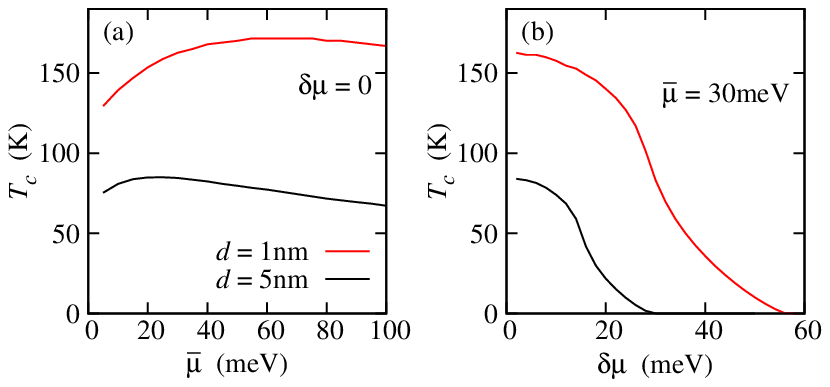}\\%
	\includegraphics[]{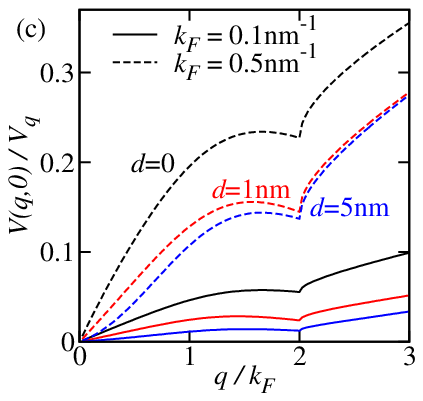}%
	\includegraphics[]{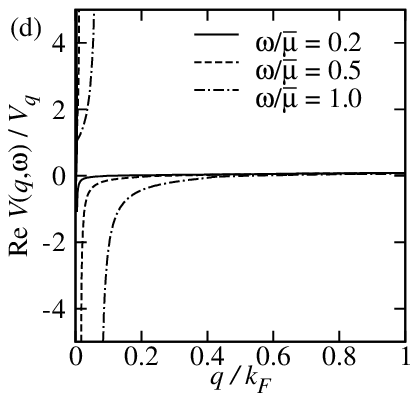}
	\caption{(Color online.)
	(a) $T_c$ as a function of \mubar with $\dmu=0$ for double
	bilayer graphene with the unscreened interlayer interaction.
	(b) $T_c$ as a function of $\dmu$ with $\mubar = 30\meV$.
	(c) The statically screened interlayer interaction for DQBG.
	(d) The dynamically screened interlayer interaction for DQBG in the
	high density regime ($k_F = 0.2\nm^{-1}$).
	\label{fig:bilayer}}
\end{figure}

We first show results for the unscreened interaction, using $V(q) = 2\pi
e^2 e^{-qd}/(\epsilon q)$. In Fig.~\ref{fig:bilayer}(a) we show $T_c$ as a
function of \mubar and find that it is of the same order as for the
equivalent case in DMG, i.e. $T_c \sim 100\mathrm{K}$ for realistic
interlayer separation [c.f. Fig.~\ref{fig:Tc}(c)].
However, the non-monotonicity of $T_c$ as a function of overall density
is unlike the monolayer case indicating that the interlayer separation
and density of electrons have a more complex relationship than in the
linear spectrum. This is due to the effective interaction parameter
$r_s$ being constant for monolayer graphene, but decreasing as
$1/\sqrt{n}$ as a function of density for bilayer graphene. Therefore,
as density increases, the reduced strength of the interactions in DQBG
manifests as a smaller excitonic gap, thus reducing $T_c$. 
Figure \ref{fig:bilayer}(b) shows $T_c$ as a function of \dmu for
$\mubar=30\meV$. This shows
qualitatively identical behavior as for DMG, indicating that within our
BCS mean-field theory, the details of the underlying band structure do
not qualitatively affect the response of the excitonic superfluid to
asymmetrical layer doping.
For comparison with experimental data where carrier density is a more
useful variable, we note that for bilayer graphene in the
single-particle limit that $\bar{n}= 2 \mast \mubar /(\pi\hbar^2)$ and
so the fluctuations are linear and $\delta n = 2 \mast \dmu / (\pi
\hbar^2)$.

As shown in Ref.~\onlinecite{hwang-prl101}, the static screening
for quadratic bilayer graphene is somewhat stronger than for
monolayer graphene, not just because the dimensionless polarizability is
larger, but because the density of states in the prefactor is also
larger at low and moderate doping.
This indicates that the interlayer interaction should be weaker in DQBG
compared with DMG for comparable parameters. 
We find that the size of the gap is  smaller than the accuracy of
the numerical procedure that we employ. This indicates that,
within our approximations for the statically screened interaction,
$T_c$ is expected to be smaller than a fraction of $1\mathrm{nK}$.
In contrast to the DMG case, the interlayer interaction $V(q,0)$ given
in Eq.~\eqref{eq:Vqomega} is not a universal function of $q/k_F$ for
DQBG in the $k_F d=0$ limit. 
Increasing the electronic density decreases the efficiency of the
screening and allows the interlayer interaction to be stronger. 
This is a substantial qualitative difference
between the screening in monolayer and bilayer graphene, which is
understandable since quadratic bilayer has a constant density of states
in contrast to the linear-in-energy density of states of monolayer
graphene.
The kink at $q=2k_F$ corresponds to the $2k_F$ anomaly for $\Pi(q)$.
\cite{hwang-prl101} 
Also, increasing $d$ will reduce the overall interaction strength which 
allows the screening to be more efficient,
reducing $V(q,0)$ with respect to $V_q$, as in DMG.

For dynamic screening, we use previous results \cite{sensarma-prb82} for
the finite frequency polarizability to determine the interaction
potential. Figure \ref{fig:bilayer}(d) shows the dynamically screened
interaction in the high density regime, where the interaction is the
strongest. The high-$q$ limit is the same as the static screening case,
and we find that the dynamic screening reaches this limit even faster
than in DMG. Therefore, the interaction strength is weaker in DQBG and
$T_c$ is suppressed. 

We now briefly comment on the role of inhomogeneity in DQBG.
Since it is known that the response of the charge distribution to
charged impurities in bilayer graphene is qualitatively similar to that
in monolayer graphene,\cite{rossi-prl101, rossi-prl107} it is very
likely that the presence of charged impurities in the environment of the
DQBG will have a similar detrimental effect on the stability of the
condensate as for DMG.
However, it is also known that bilayer graphene is, in general, somewhat
more robust against ripples and corrogations than monolayer
graphene,\cite{chang-epjb78}
indicating that charge inhomogeneity generated by this form of disorder
may be slightly less important.

\section{Conclusion}

In summary, we have presented a comprehensive analysis of the effect of
charge inhomogeneity on $T_c$ for the excitonic superfluid in double
layer graphene systems.
We find that the existence of charge inhomogeneity and ripples is
likely to be the limiting factor for the stability of the condensate,
but that the cleanest samples at low temperature should allow for the
detection of the condensate.
If the graphene layers are suspended, particular care must be taken to
ensure minimal rippling since even in the absence of charged
impurties, this may be a significant source of density inhomogeneity.
We also investigated the equivalent situation in DQBG, showing that in
the unscreened case, the quadratic nature of the low energy bands does
change the qualitative behavior of $T_c$ as a function of \mubar, but
that the maximum achievable $T_c$ is quite similar in both systems.
However, static screening is somewhat stronger in DQBG
but the details of the realistic screening are still unknown.
Our most important conclusion is that the only hope for achieving
excitonic condensation in graphene is to use very flat, ultra-pure 
graphene with
very low density of charged impurities so that the induced charge
inhomogeneity is minimized.  
Even then, if the operative interlayer pairing interaction turns out to
be the statically screened Coulomb interaction, there is very little
hope for the observation of the interlayer superfluid state at any
reasonable temperatures.

\acknowledgments
DSLA and SDS are supported by US-ONR, NSF-JQI-PFC and LPS-CMTC.
MRV and ER are supported by ONR-N00014-13-1-0321 and the Jeffress
Memorial Trust. 
Some of the calculations were carried out on the SciClone Cluster at
College of William and Mary.

\appendix

\section{Screening \label{app:screening}}

\begin{figure}
	\centering
	\includegraphics[width=8.5cm]{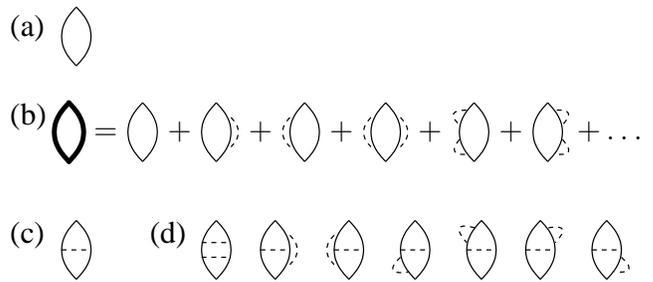}
	\caption{(a) The diagram for $\Pi(\vq,\omega)$ in our
	calculation.
	(b) The diagram for $\Pi(\vq,\omega)$ in the calculation of
	Refs.~\onlinecite{lozovik-prb86, perali-prl110}.
	(c) The first-order diagram absent from (b).
	(d) The second-order diagrams absent from (b).
	The dashed lines represent interlayer electron--electron
	interactions.  
	\label{fig:diagrams}}
\end{figure}

In this appendix, we clarify certain aspects of the screening models
that are used both in the literature and in our own calculation.
In particular, Refs.~\onlinecite{sodemann-prb85, perali-prl110,
lozovik-prb86} use what the authors describe as a ``self-consistent
screening approximation''.
In this theory, the gap at the Fermi energy generated by the superfluid
pairing is self-consistently included in the screening, with the result
that the interlayer interaction is increased dramatically, enabling the
condensate to be stable at temperatures of the order of $100\mathrm{K}$.
The fundamental issue we wish to raise is that the expansion of the
polarization function on which these works rely does not include
all diagrams of any arbitrary order in the electron--electron
interaction. In Fig.~\ref{fig:diagrams}(a), we show the zeroeth order
expansion which we use in our calculation, and in
Fig.~\ref{fig:diagrams}(b) is the diagram and expansion to second order
used in Refs.~\onlinecite{perali-prl110, lozovik-prb86}.
Figures \ref{fig:diagrams}(c) and \ref{fig:diagrams}(d) show,
respectively, the first order and second order diagrams that are not
present in this expansion. 
The absence of these diagrams violates the Ward identity for
current conservation.
As described in great detail in Refs.~\onlinecite{fertig-prl65,
fertig-prb44}, the theoretical formulation of the dielectric
response of a superconductor in the gapped symmetry-broken phase is a
formidable task, which cannot be simulated simply by incorporating the
self-energy in the polarization bubble diagram since such an
approximation is not conserving as it leaves out many other diagrams
(see Fig.~\ref{fig:diagrams} and Refs.~\onlinecite{fertig-prl65,
fertig-prb44}) in each order.
It is therefore theoretically more meaningful to use a clearly defined
perturbative approach where the screened interaction is calculated in
the normal state (as done in the current paper) and then the BCS mean
field theory is carried out on this normal state interaction function.
Our theory thus corresponds to the leading-order conserving perturbative
approximation where only the bare bubble [Fig.~\ref{fig:diagrams}(a)] is
used for the polarizability, thus satisfying the Ward identities.
While this may not be a quantitatively accurate approximation,
it is guaranteed to satisfy conservation laws. The inclusion of the
vertex corrections to the ladder diagrams, which for the systems studied
are not captured by the Migdal theorem, would be desirable but is well
beyond the scope of this work (and, for that matter, all existing work
in this subject) whose focus is the effect of long-range
disorder on the conditions, in particular $T_c$, for the realization of
excitonic superfluid states in DMG and DQBG.
The issue of a better screening approximation going beyond RPA remains
open for the graphene interlayer superfluidity problem since such a
theory must combine the conserving approximation of
Refs.~\onlinecite{fertig-prl65, fertig-prb44} with the peculiar band
structure and chirality of graphene physics.

In the current manuscript, we have demonstrated that DMG and DQBG have
$T_c$ of the same order for the equivalent level of approximation for
the interlayer screening.
How the consistent approximation of Refs.~\onlinecite{fertig-prl65,
fertig-prb44} would affect the excitonic condensation is currently
unknown and remains an interesting open question, but it is not
unreasonable to assume that the DMG and DQBG systems would still have
$T_c$ of the same order in this case.

\bibliography{bibtex-sorted}

\end{document}